\documentclass[showpacs,amsmath,amssymb,aps,showkeys,floatfix,prc,onecolumn,a4paper,preprint,nofootinbib]{revtex4-2}
\usepackage{graphicx}
\usepackage{epstopdf}
\usepackage{dcolumn}
\usepackage{bm}
\usepackage{epsfig}
\usepackage{amsfonts}
\usepackage{amssymb,amscd}
\usepackage{hyperref}
\usepackage{xcolor}
\hypersetup{
    colorlinks=true,                
    breaklinks=true,                
    urlcolor= black,                
    linkcolor= blue,                
    bookmarksopen=false,
    filecolor=black,
    citecolor=blue,
    linkbordercolor=blue,
    pdfborder = {0 1 0}
}

\begin{document}

\title{Double charmed meson production in $pp$ and $pA$ collisions  at the LHC within the dipole approach in momentum representation}


\author{G. Sampaio dos Santos}
\affiliation{High Energy Physics Phenomenology Group, GFPAE  IF-UFRGS \\
Caixa Postal 15051, CEP 91501-970, Porto Alegre, RS, Brazil}

\author{G. Gil da Silveira}
%\affiliation{CERN, PH Department, 1211 Geneva, Switzerland}
\affiliation{High Energy Physics Phenomenology Group, GFPAE  IF-UFRGS \\
Caixa Postal 15051, CEP 91501-970, Porto Alegre, RS, Brazil}
%\affiliation{Departamento de F\'{\i}sica Nuclear e de Altas Energias, Universidade do Estado do Rio de Janeiro\\
%CEP 20550-013, Rio de Janeiro, RJ, Brazil}

\author{M. V. T. Machado}
\affiliation{High Energy Physics Phenomenology Group, GFPAE  IF-UFRGS \\
Caixa Postal 15051, CEP 91501-970, Porto Alegre, RS, Brazil}

\begin{abstract}
A study of double charmed meson production in proton-proton and proton-nucleus collisions at the LHC energies is performed.
Based on the color dipole formalism developed in the transverse momentum representation and the double parton scattering mechanism, predictions are made for the transverse momentum differential cross section for different pairs of $D$-mesons. The theoretical results consider the center-of-mass energy and forward rapidities associated to the measurements by the LHCb Collaboration. The results considering different unintegrated gluon distributions are presented and compared to data and predictions for proton-nucleus collisions are provided.
\end{abstract}

\maketitle

\section{Introduction}
\label{intro}

The heavy quark production, especially at high energies, can provide access to particular kinematic regions that allow investigating the perturbative Quantum Chromodynamics (pQCD) regime \cite{godwiss,andronic}. The heavy quark mass is employed as a hard factorization scale and consequently pQCD calculations can be performed \cite{nason,mangano,cacciari,frixione,anderle} as a valid description. Furthermore, heavy quarks measurements can be used to extract nonperturbative information on the heavy flavor fragmentation functions \cite{anderle}. Recent experimental measurements of charmed mesons, particularly $D$-mesons \cite{LHCb5pp,LHCb13,ALICE,ALICE1,LHCb5,LHCb816,ALICE502}, covering a wide range of values for center of mass energies, transverse momentum, and rapidity are available in the literature. In a hadronic collision at the Large Hadron Collider (LHC), heavy quarks are produced via the hard scattering between the parton constituents of the incident hadrons and the $D$-mesons measured in the final state are formed by the hadronization process. At high energies typical of the LHC, it is relevant to study the physics associated to the small values of the Bjorken variable $x$, where one expects that the nonlinear effects of the QCD play a significant role on the description of the observables.  
 
Viability studies can gain new prospects since a new kinematic domain has been probed at the LHC due its high energy and luminosity together with the measurements at high precision achieved by its experiments. As the energy increases along with the high density of partons in the hadron wave function, the probability of occurrence of multiple parton interactions (MPIs) is enhanced. As an effect, double parton scattering (DPS) processes may play an important role in the production mechanisms as indicated by both theoretical and experimental investigations in the literature. Moreover, the ratio concerning the probabilities of DPS to SPS grows in energy \cite{paver, mekhfi, sjostrand} and DPS contribution can not be disregarded. 
From the experimental scenario, it has been demonstrated at the LHC energies that the DPS contribution in proton-proton ($pp$) collisions is similar to the SPS \cite{luszczak, cazaroto}.
The DPS picture consists on two quarks/gluons interacting with other partons in an independent form within the same reaction. Thus, MPI of some particular states in hadronic collisions emerge as a consequence of the DPS processes. It is already known that charmonium and open charm mesons production present relatively large cross sections at high energies and they can be used to investigate the SPS and DPS processes. There are many studies considering the DPS mechanism as a source of the quarkonium production in the double charmonium as well as charmonium plus open charm production \citep{brodsky}.
Furthermore, the DPS accounts for a fundamental class of processes that allows the study of the spatial structure of hadrons \cite{diehl}, heavy quark-antiquark asymmetries \cite{gauld}, parton-parton correlations in the nucleon wave function \cite{blok, ostapchenko}, and the double parton distribution functions (DPDFs) \cite{rinaldi, rinaldi1, diehl1}. In this work, we provide predictions for double $D$-meson production considering the theoretical framework of the color dipole approach. Our results are directly compared to the measurements performed in the high-energy kinematic regime accessible in $pp$ and proton-nucleus ($pA$) collisions at the LHC.

Based on theoretical scenarios, the $D$-meson production cross section is obtained in the framework of the QCD calculations performed within the collinear factorization \cite{collins} or the $k_{T}$-factorization approach \cite{gribov, march, levin, catani}. In the latter, the $D$-meson hadroproduction is described in terms of the gluon densities by the unintegrated gluon distribution (UGD). Also, it considers the transverse momenta of the initial partons and has a dependence on their momentum fraction $x$ and the factorization scale $\mu_{F}$. The UGDs has to be parameterized, where the associated models are based on different underlying physical assumptions, concerning particularly their rapidity dependence, $Y = \mathrm{ln}(1/x)$, and transverse momentum, $k_{\perp}$. In the present work, the color dipole formalism \cite{nik,rauf} will be applied to the heavy quark production, which is a suitable framework in describing the phenomenology associated to different processes at small-$x$ and currently employed in many studies. The respective dipole amplitude is associated to the dipole transverse momentum distribution (TMD), namely, the intrinsic dipole $k_{\perp}$-distribution. In the limit of large transverse momentum, the dipole TMD is approximately equivalent to the UGD. 

Here, TMDs models  based on gluon saturation approach will be used. The nonlinear gluon QCD effect is associated with a transition region limited by a $x$-dependent saturation scale, $Q_s(x)$, the transverse momentum scale that marks the onset of the gluon saturation physics. This nonlinear QCD phenomenon is expected to occur at the low-$x$ kinematic region where the gluon recombination process is well established. High energy measurements of the double $D$-meson distribution allow to investigate this dense and saturated regime. The calculations will be based on the SPS cross section for $D$-meson production through the color dipole formalism in transverse momentum representation presented in Refs.~\cite{gsds,gsds1}. Another important aspect is the addition of nuclear effects to investigate the $D$-pair production in $pA$ collisions. The QCD dynamics at low $x$ and high gluon densities \cite{salgado} can be probed in $pA$ collisions, which serve as a baseline for studies in nucleus-nucleus ($AA$) collisions. Moreover, the high energy description given by the Color Glass Condensate (CGC) effective theory \cite{tribedy,albacete,amir} assumes that the nucleus is a saturated gluonic system. Hence, we expected that the DPS mechanism in $pA$ collisions is enhanced in relation to the $pp$ mode, since there is a possibility of the proton to be scattered from two or more different nucleons inside the nucleus \cite{bc}, and the corresponding enhancement factor has been estimated to be approximately 3 considering proton-lead ($p$Pb) collisions \cite{luszczak,cazaroto1,david,helenius}. Furthermore, the DPS production can be used as a sensitive tool in view of constraining the nuclear PDF (nPDF) in $pA$ collisions, taking into account a dependence on position in the ion \cite{shao}.

The paper is organized as follows: In Sec.~\ref{form} the basic assumptions and expressions regarding the theoretical formalism for obtaining the double $D$-meson production cross section in $pp$ and $pA$ collisions are presented, together with the analytical models for the UGDs. In Sec.~\ref{res} the results are shown and compared to the experimental measurements in the forward rapidities probed by the LHCb experiment in $pp$ mode \cite{LHCbpp}. Predictions are performed for the DPS cross section in the $pA$ collisions \cite{LHCbpPb}. In Sec.~\ref{conc} we summarize our main conclusions. 

\section{Theoretical formalism}
\label{form}

We start considering a model that describes the cross section for the double $D$-meson production in a simple generic form leading to the so-called {\textit{pocket formula}}. The model is based on the assumption that the parton distribution functions (PDFs) of two partons in the same projectile are independent and then two separate partonic
interactions occur (a DPS process) generating the associated production cross section of two final-state particles. Namely, the DPS cross section is obtained by the product of the two corresponding individual SPS cross sections where the respective SPS processes are uncorrelated and do not interfere with each other \cite{cazaroto1, david}. Therefore, the DPS cross section for double $D$ meson production is given by:
\begin{eqnarray}
\sigma^{DPS}_{pp,\,pA \rightarrow D_1\,D_2} = \beta\, \frac{\sigma^{SPS}_{pp,\,pA \rightarrow D_1} \cdot \sigma^{SPS}_{pp,\,pA \rightarrow D_2}}{\sigma_{eff}^{pp,\,pA}}\,,
\label{dpsxs}
\end{eqnarray}
where $\sigma_{eff}$ is an effective cross section connected with the collision geometry \cite{rc} and is interpreted as the effective transverse overlap of the partonic interactions that configures the DPS mechanism. Assuming a geometric interpretation, the $\sigma_{eff}$ can be determined from the integral of the overlap function over the impact parameter. In the literature the parameter $\sigma_{eff}$ has been determined by using the measurements obtained in experiments at the Tevatron \cite{CDF} and the LHC \cite{LHCbpPb} for the DPS production in $p\bar{p}$ and $p$Pb collisions. The extracted values presented in those studies are $\sigma_{eff}^{pp} = 14.5 \pm 1.7$~mb and $\sigma_{eff}^{pPb} = 4.3 \pm 0.5$~b, respectively. Additionally, the quantity $\beta$ in Eq.~(\ref{dpsxs}) accounts for the different configurations of the final state. Explicitly, $\beta = 1/4$, if $D_1$ and $D_2$ are identical and non-self-conjugate, $\beta = 1$, if $D_1$ and $D_2$ are different and either $D_1$ or $D_2$ is self-conjugate, and $\beta = 1/2$ otherwise \cite{LHCbpp}. 

At this level, the color dipole formalism resums all orders (diagrammatic contributions) in log(1/$x$) and the higher twist contributions to inclusive observables. Hence, the corresponding contributions are taken into account in order to evaluate the observables, for instance, the double $D$-meson production. Nevertheless, it is not clear if the NLO corrections within collinear factorization approach are all accounted for in the dipole formalism, which are relevant for quarkonium production. Studies of $J/\psi$-pair production have considered the addition of different production mechanisms in distinct kinematic regions, and
both DPS and the NLO QCD corrections to SPS are crucial to account for the existing data. In Ref.~\cite{ls} the DPS yields are computed by using NLO contributions, with a large enhancement seen between the yields at LO and NLO. Thus, improvements are needed to the formalism if one has to account for the NLO contributions.

In this work the QCD dipole framework will be used to compute the SPS cross section for heavy meson production. This development assumes the target rest-frame and that the production process can be determined by a color dipole, $Q\bar{Q}$, that interacts with the color field of the proton/nucleus. The $D$-meson production is described by the cross section of the process $g+p(A) \rightarrow Q\bar{Q}+X$, where the corresponding $Q\bar{Q}$ comes from a virtual gluon fluctuation, produced in singlet or color-octet states. In the momentum representation the heavy quark transverse momentum distribution can be obtained in terms of the gluon dipole TMD, ${\cal T}_{\mathrm{dip}}$ \cite{vic}, in the following way: 
\begin{eqnarray} 
\frac{d^3\sigma{(gp \rightarrow Q\bar{Q}X)}}{d\alpha d^2 p_T } &=& 
\frac{1}{6\pi} \int \frac{d^2 \kappa_{\perp}}{\kappa^{4}_{\perp}}  \alpha_s(\mu_{F}^2)\, {\cal T}_{\mathrm{dip}}(x_2,\kappa^{2}_{\perp})\,\bigg\{\bigg[\frac{9}{8}{\cal{I}}_0(\alpha,\bar{\alpha},p_T) - \frac{9}{4} {\cal{I}}_1(\alpha,\bar{\alpha},\vec{p}_T,\vec{\kappa}_{\perp}) \nonumber \\ 
&+& {\cal{I}}_2(\alpha,\bar{\alpha},\vec{p}_T,\vec{\kappa}_{\perp}) + \frac{1}{8}{\cal{I}}_3(\alpha,\bar{\alpha},\vec{p}_T,\vec{\kappa}_{\perp})\bigg] + \left[\alpha \longleftrightarrow \bar{\alpha}\right]\bigg\} \,.  
\label{proxs} 
\end{eqnarray}
In Eq.~(\ref{proxs}), $\alpha_s(\mu_{F}^2)$ represents the running coupling at one-loop level dependent on the scale $\mu_{F}^2 = M^{2}_{Q\bar{Q}}$, being $M_{Q\bar{Q}}$ the invariant mass of the $Q\bar{Q}$ pair determined by the mass ($m_{Q}$) and transverse momentum ($p_T$) of the heavy quark, $M_{Q\bar{Q}}\simeq 2\sqrt{m_{Q}^2+p_T^2}$. Moreover, $\alpha$ and $\bar{\alpha} = 1 - \alpha$ are the gluon momentum fractions exchanged with the heavy quark and antiquark, respectively. 
In addition, Eq.~(\ref{proxs}) includes the auxiliary quantities ${\cal{I}}_i$ ($i=0,1,2,3$), which depends on the longitudinal momentum fractions $\alpha/\bar{\alpha}$, quark transverse momentum $p_T$, and gluon transverse momentum $\kappa_{\perp}$. Expressions for these quantities are given in Refs.~\cite{vic,gsds,gsds1}.

The intrinsic dipole TMD is approximately equal to the UGD function times $\alpha_s$ \cite{gbw,bart,albcgc,altcgc}, when the momentum of the gluon in the target is sufficiently large, such that $\kappa_\perp\gg\Lambda_{\rm QCD}$. This assumption implies that a relation between the $k_\perp$-factorization and the dipole approach can be established, $
{\cal T}_{\rm{dip}}(x_2,\kappa_{\perp}^2)\simeq\alpha_s\,{\cal F}(x_2,\kappa_{\perp}^2)$, with ${\cal F}$ denoting the target UGD. The ${\cal T}_{\rm{dip}}$ is connected to the dipole cross section $\sigma_{q\bar{q}}$, since one is able to extract the respective TMD for a particular dipole cross section model by applying a specific Fourier transform (see Refs.~\cite{bart,nik1}). Furthermore, considering the $k_{\perp}$-factorization formalism and disregarding the primordial gluon momentum, the gluon UGD, ${\cal F}(x,k_{\perp}^2)$, and the collinear gluon distribution, $g(x,Q^2)$, are related by
\begin{eqnarray}
g(x_1,\mu_{F}^2) = \int^{\mu^2_F}dk_{\perp}^2{\cal{F}}(x_1,k_{\perp}^2)\,,
\label{fg}
\end{eqnarray}
where $x_{1}$($x_{2}$) is the fractional longitudinal momentum of the projectile (target) as function of the heavy quark pair rapidity $y$, namely, $x_{1,2} = \frac{M_{Q\bar{Q}}}{\sqrt{s}}\, e^{\pm y}$, where $\sqrt{s}$ stands for the collision center of mass energy.

The UGD can not be computed by first principles, hence a number of parameterizations are available. Here, we will consider the analytical models for the UGD in protons provided in Refs.~\cite{gbw,mpm,ww}. Two of them present geometric scaling property  \cite{Stasto:2000er,Beuf:2008mf,Praszalowicz:2013iyi,munier}, meaning that the UGD depends on the ratio $\tau = k_{\perp}^2/Q_{s}^2(x)$ instead of depending separately on $x$ and  $k_{\perp}$. The first one is the gluon UGD from the Golec-Biernat and W\"usthoff (GBW) saturation model \cite{gbw} which reads,
\begin{eqnarray}
{\cal F}_{GBW}(x,k_{\perp}^2)=\frac{3\,\sigma_{0}}{4 \pi^2\alpha_{s}} \tau\,\mathrm{exp}\left(-\tau\right) \,,
\label{FGBW}
\end{eqnarray}
where $\alpha_{s} = 0.2$ and $Q_{s}^2(x) = (x_0/x)^{\lambda}\,\rm{GeV}^2$ is the proton saturation scale, with the following set of parameters: $\sigma_{0} = 27.43$~mb, $x_{0} = 0.40 \times 10^{-4}$, and $\lambda = 0.248$, extracted from the fit to proton structure function, $F_2^p$, data at small-$x$ and reported in Ref.~\cite{gbwfit}.

The second analytical model is the Moriggi-Peccini-Machado (MPM) parametrization \cite{mpm}, which is based on geometric scaling property. This model reproduces correctly the hadron spectrum in $pp$ collisions at high energies. The MPM parameterization is given by, 
\begin{eqnarray}
{\cal F}_{MPM}(x, k_{\perp}^2)=\frac{3\,\sigma_{0}}{4\pi^2\alpha_{s}}\,
\frac{\tau\,[1+\varepsilon (\tau)]}{\left(1+\tau \right)^{2+\varepsilon (\tau)}}\,,
\label{FMPM}
\end{eqnarray}
where $\alpha_{s} = 0.2$ and the saturation scale has the same form as the GBW model, however a fixed $\lambda = 0.33$. The scaling variable is denoted by $\tau$ as before. The power-like behavior of the spectrum of gluons at high momentum is defined in terms of the function $\varepsilon (\tau) = a\tau^{b}$.
The following set of parameters is determined by fitting DIS data available at low-$x$ in Ref.~\cite{mpm}: $\sigma_{0} = 19.75$~mb, $x_{0} = 5.05 \times 10^{-5}$, $a = 0.075$ and $b = 0.188$.  

The third analytical parametrization was proposed in Ref.~\cite{ww}, labeled here as WW UGD. This model is inspired by the method of virtual quanta proposed by Weizs\"acker and Williams (WW), considering the hard gluon TMD with the asymptotic behavior of one gluon exchange at large gluon transverse momenta between a point-like parton and a hard probe. This gluon exchange behaves like a virtual photon exchange, then the associated virtual gluon density resembles the WW virtual photon density around a point-like charge. In this parametrization, the WW UGD is given by  
\begin{equation}\label{FWW}
{\cal F}_{WW}(x,k_{\perp}^2) = (N_1/k_0 ^2)(1-x)^7\times
\begin{cases}
 (x^{\lambda} k_{\perp}^2 / k_0^2)^{-b} &  k_{\perp}^2 \geq k_0^2, \\
 x^{-\lambda b} & k_{\perp}^2 < k_0^2,
\end{cases}
\end{equation}
where the normalization constant is $N_1 = 0.6$, $k_0 = 1$~GeV, and $\lambda = 0.29$. The factor $(1 - x)^7$ is introduced to account for the suppression of the gluon distribution at large $x$ whereas the phenomenological parameter $b$ controls the $k_{\perp}$-scaling of the gluon distribution. It has been shown in  Ref.~\cite{ww} that the shape of WW TMD is essential in order to obtain the correct description of the Lam-Tung relation breaking at the $Z^0$ hadroproduction in the context of $k_{\perp}$-factorization formalism.

The hadronic cross section of the process $pp \rightarrow Q\bar{Q}X$ is given by the convolution between the $gp \rightarrow Q\bar{Q}X$ cross section and the projectile gluon UGD,
\begin{eqnarray} 
\frac{d^{4}\sigma{(pp \rightarrow Q\bar{Q}X)}}{dy d\alpha d^2p_T} = g(x_1,\mu_{F}^2)\,
\frac{d^{3}\sigma{(gp \rightarrow Q\bar{Q}X)}}{d\alpha d^2p_T} \,,
\label{hdeq}
\end{eqnarray}
where intrinsic transverse momentum of partons in the projectile has been disregarded. As a matter of self-consistency, $g(x_1,\mu_{F}^2)$ will be obtained by using Eq.~(\ref{fg}) and considering the the same UGD considered for the target. 

Furthermore, in order to investigate the $D$-meson production, a hadronization process of the heavy quarks accounting for the probability that a heavy quark fragments into a meson is required. As a result, the $D$-meson production spectrum is obtained by a convolution of the heavy quark cross section and the fragmentation function,
\begin{eqnarray} 
\frac{d^{3}\sigma{(pp \rightarrow DX)}}{dY d^2P_T} = \int_{z_{\mathrm{min}}}^1 \frac{dz}{z^2} 
D_{Q/D} (z,\mu_{F}^2) \int_{\alpha_{\mathrm{min}}}^1 d\alpha \frac{d^{4}\sigma{(pp \rightarrow Q\bar{Q}X)}}{dyd\alpha d^2p_T} \,,
\label{Dmes}
\end{eqnarray}
with $z$ being the fractional transverse momentum of the heavy quark carried by the $D$-meson and $D_{Q/D}(z,\mu_{F}^2)$ denotes the meson fragmentation function. The Kneesch-Kniehl-Kramer-Schienbein parameterization (KKKS) \cite{kkks08} will be employed in the numerical calculations. The $D$-meson mass and rapidity are $m_D$ and $Y=y$, respectively. The transverse momentum of the $D$-meson is represented by $P_T$ and is related to the quark transverse momenta in the form $p_{T} = P_{T}/z$. Finally, in Eq.~(\ref{Dmes}), the lower limits of integration over $z$ and $\alpha$ are expressed by $z_{\mathrm{min}}= (\sqrt{m_{D}^2 + P_{T}^2}/\sqrt{s}) e^{Y}$ and $\alpha_{\mathrm{min}}=(z_{\mathrm{min}}/z)\sqrt{(m_{Q}^2 z^2 + P_{T}^2)/m_{D}^2 + P_{T}^2}$, respectively.

As far the meson production in $pA$ collisions is concerned, the Glauber model applied to hard processes can be employed using the cross section for $pp$ collisions as a baseline as discussed before. However, we investigate in this work a different approach by using a nuclear UGD replacing the proton one. This is related to the evaluation of the dipole-nucleus amplitude, $N_A(x,r)$,  and the associated QCD nuclear effects that occur in high-energy collisions with heavy nuclei targets. The nuclear effects can be described within the color dipole formalism by the geometric scaling in the dipole-nucleus amplitude. The geometric scaling property derived from parton saturation models \cite{salgado1} assumes that the nuclear effects are embedded into the nuclear saturation scale, $Q_{s,A}$, and on the nucleus transverse area, $S_A=\pi R_A^2$ (with $R_A\simeq 1.12 A^{1/3}$ fm) with the proton case as reference, $S_p=\sigma_0/2=\pi R_p^2$.
Consequently, the proton saturation scale, $Q_{s,p}$, is properly replaced by the nuclear saturation scale, $N_A(x,r) =  N(rQ_{s,p}\rightarrow rQ_{s,A})$, where
\begin{eqnarray}
Q_{s,A}^2=Q_{s,p}^2\left(\frac{AS_p}{S_A}\right)^{\Delta},  \label{qs2A} 
\end{eqnarray}
with the quantities $\Delta = (0.79)^{-1}$ and $S_p=1.55$~fm$^2$ \cite{salgado1}. The geometric scaling approach is able to describe the nuclear modification factor for the nuclear structure functions, $R = F_2^A/AF_2^p$, at the small-$x$ region.

Therefore, based on the assumptions from the geometric scaling approach, one obtains a simplified expression for the $pA$ cross section given an UGD for protons which presents scaling. This is the case for the GBW and MPM parametrizations. Namely, the scaling is translated into the cross section for the $D$-meson production in $pA$ collisions in the following way,
\begin{eqnarray}
\frac{d^{3}\sigma{(pA \rightarrow DX)}}{dY d^2P_T} = \left(\frac{S_A}{S_p}\right) \left.\frac{d^{3}\sigma{(pp \rightarrow DX)}}{dY d^2P_T}\right|_{Q_{s,p}^2(x_2)\rightarrow Q_{s,A}^2(x_2)}\,.
\label{prescr}
\end{eqnarray}
The approach reported above has been used in the literature, for instance in Refs.~\cite{gsds2,gsds3,gsds4} in studies regarding the prompt photon production in $pA/AA$ collisions. 

As presented in Ref.~\cite{armesto}, another possibility is to obtain the nuclear UGD by using the Glauber-Gribov approach for the dipole-nucleus cross section with the GBW model as input. The advantage is that such parameterization contains the dependence on the impact parameter. In this approach, the UGD for the nucleus one reads as \cite{armesto, armesto2},
\begin{eqnarray}
{\cal{F}}_{nuc}(x,k_{\perp},b) = \frac{3}{\pi^2\alpha_{s}}\frac{k^{2}_{\perp}}{Q_{s,p}^2} \sum_{n=1}^{\infty} \frac{(-B)^{n}}{n!} \sum_{\ell=0}^{n} C_{n}^{\ell}\frac{(-1)^\ell}{\ell} \,\mathrm{exp}\left(-\frac{k_{\perp}^2}{\ell\,Q_{s,p}^2}\right)\,,
\label{ugdfnuc}
\end{eqnarray}
with $B = AT_{A}(b)\sigma_{0}/2$ and $T_{A}(b)$ is the nuclear thickness function. The series is rapidly convergent for large nucleus and in the numerical calculation using Eq.~(\ref{proxs}) one has that ${\cal{F}}_A(x_2,k_{\perp}^2) = \int d^2b\, {\cal{F}}_{nuc}(x_2,k_{\perp},b) $. Hereafter, ${\cal{F}}_A$ will be labeled by \emph{UGDnuc}.

In the next section we will study the implications of the DPS processes in the simultaneous production of two charmed hadrons in $pp$ and $pA$ collisions at the LHC. The focus will be on the transverse momentum distributions. 

\section{Results and discussions}
\label{res}

The present study takes into account the experimental measurements of the double $D$-meson production covered by the kinematic regime
available at the LHCb experiment considering $pp$ and $pA$ collisions \cite{LHCbpPb,LHCbpp}. The corresponding predictions are obtained with the DPS mechanism within the color dipole approach in transverse momentum framework in conjunction with three distinct UGDs: GBW, MPM, and WW models. For the nuclear case, the results are calculated with the nuclear UGD labeled as \emph{UGDnuc} and also by applying the geometric scaling considering the UGD MPM denoted as \emph{GS (MPM)}.

\begin{figure*}[!t]
\begin{tabular}{cc}
\includegraphics[width=0.5\textwidth]{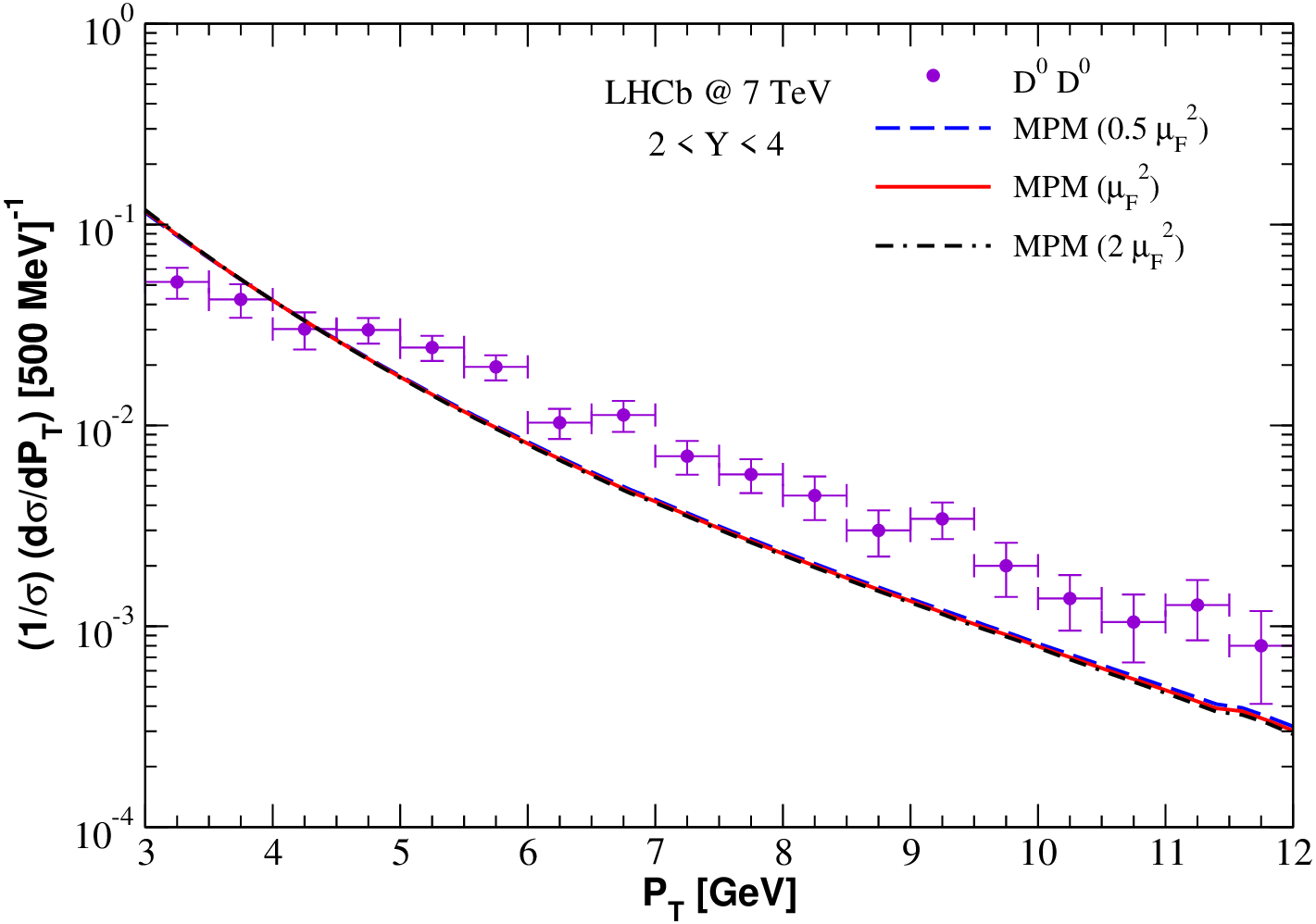} & \includegraphics[width=0.5\textwidth]{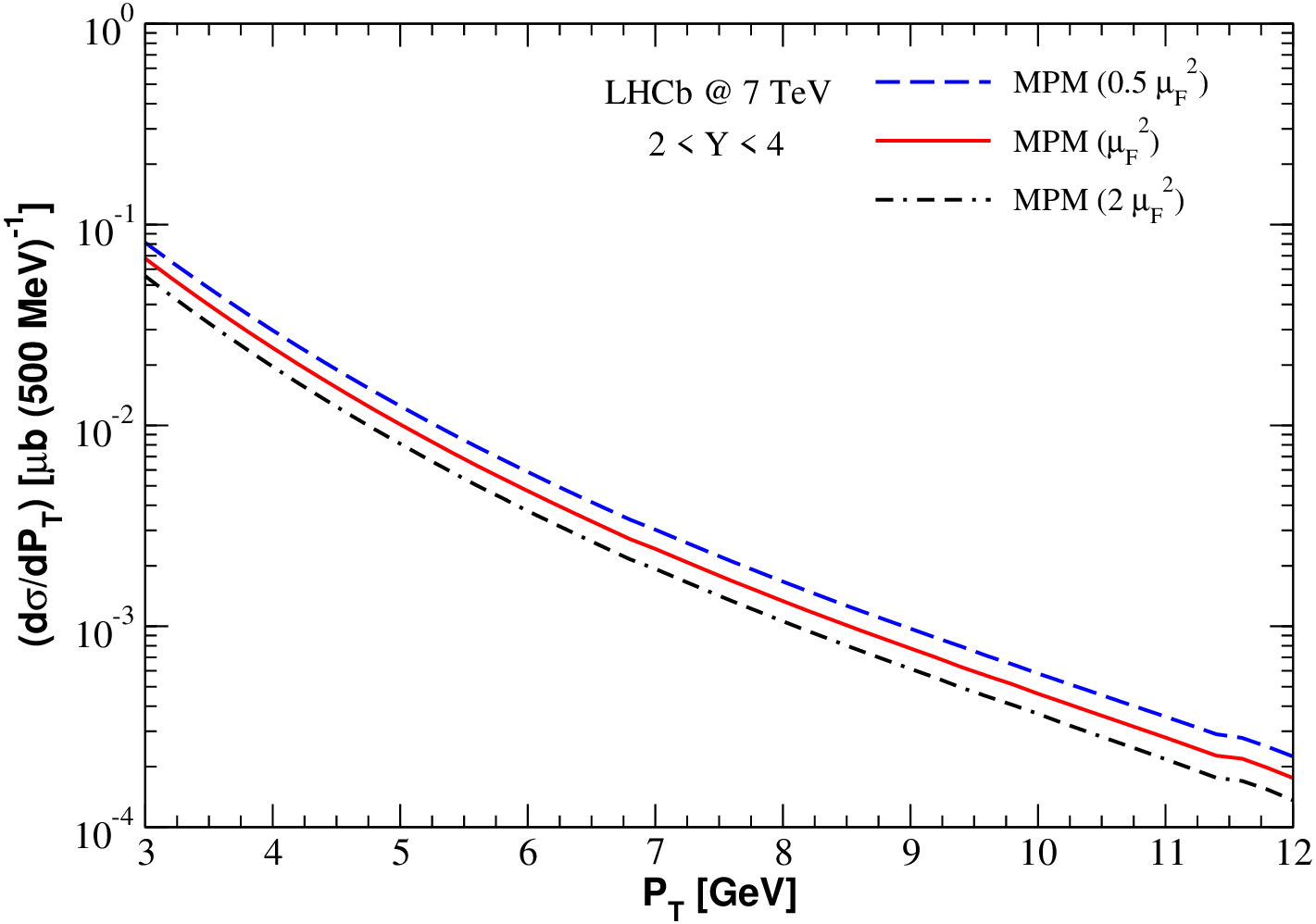} 
\end{tabular}
\caption{The uncertainties regarding the choice of the factorization scale. The results with (left panel) and without (right panel) the normalization by the integrated cross section. $D^0D^0$ measurements in $pp$ collisions from the LHCb Collaboration \cite{LHCbpp} are shown at $\sqrt{s} = 7$~TeV and  forward rapidities, $2 < Y < 4$.}
\label{pp5_fac}
\end{figure*}

\begin{figure*}[!t]
\begin{tabular}{c}
\includegraphics[width=0.75\textwidth]{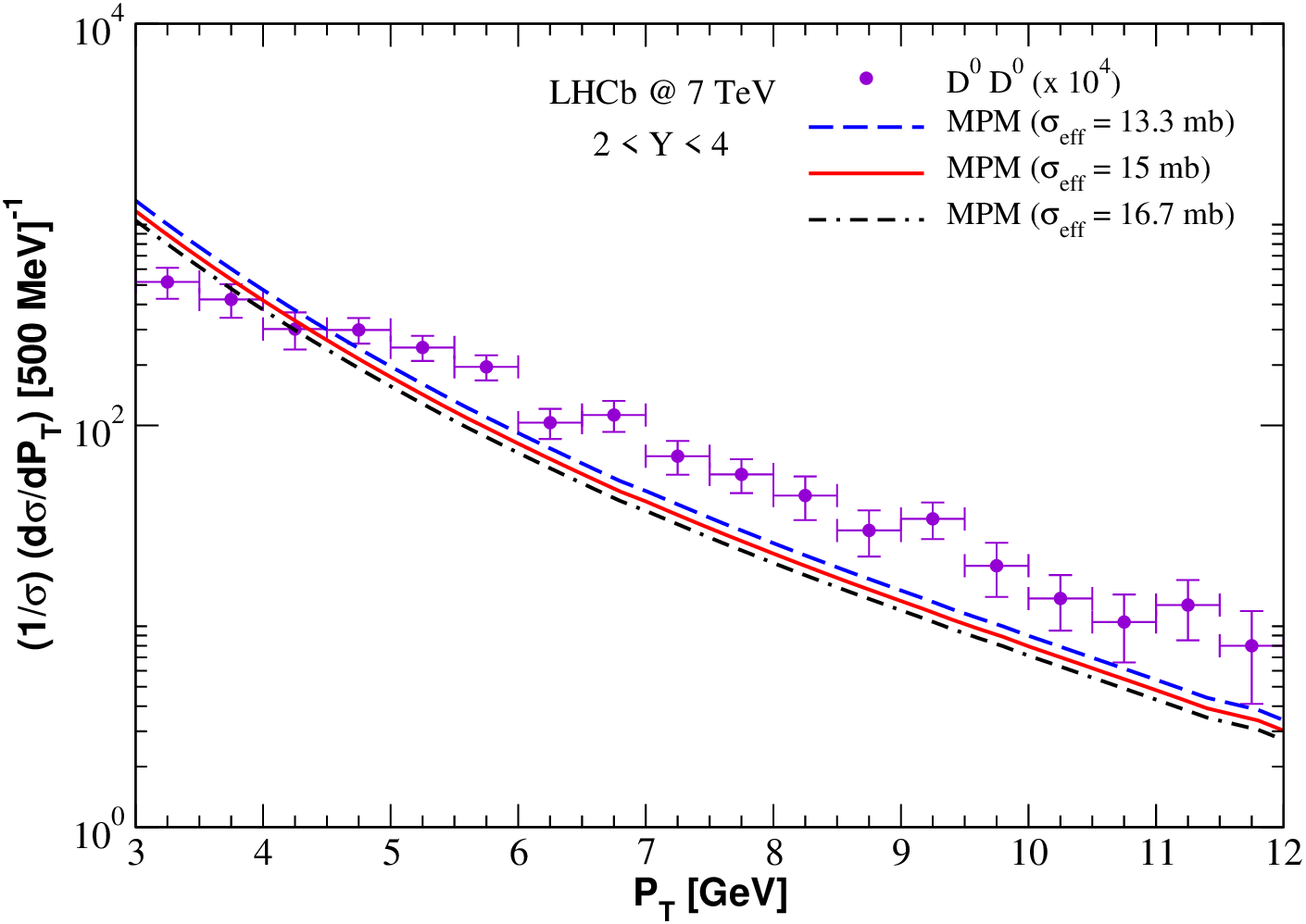} 
\end{tabular}
\caption{The uncertainties regarding the choice of the effective cross section, $\sigma_{eff}^{pp}$. The results with MPM model corresponding to the $D^0D^0$ measurements in $pp$ collisions from the LHCb Collaboration \cite{LHCbpp} at $\sqrt{s} = 7$~TeV and  forward rapidities, $2 < Y < 4$.}
\label{pp7_eff}
\end{figure*}

First, we investigate the possible sources of uncertainties in the theoretical calculations. For $pp$ collisions we have considered $\sigma_{eff}^{pp} = 15$~mb. The perturbative uncertainty associated to the factorization scale has been investigated. We consider the $D^0D^0$ production at the energy of $\sqrt{s} = 7$~TeV and rapidity bin $2 < Y < 4$ measured at the LHCb experiment \cite{LHCbpp} in $pp$ mode. For the predictions we select the MPM model along with three values of the factorization scale given by $0.5\,\mu_F^{2}$, $\mu_F^{2}$, and $2\,\mu_F^{2}$. In Fig.~\ref{pp5_fac} (left panel) the difference among the results is not visible since the transverse momentum distributions are normalized by the corresponding integrated cross section. This is the way that the LHCb Collaboration extracted the available data. On the other hand, without the normalization as seen in Fig.~\ref{pp5_fac} (right panel), the difference concerning the results become more pronounced as the $P_T$ value increases. It was verified that there is an uncertainty of around 30\% related to the result with the central value of the factorization scale which is considered in our calculations. Complementarily, we perform an analysis in order to take into account the uncertainty on $\sigma_{eff}^{pp}$ related to the theoretical calculations assuming the MPM model. We use the uncertainty reported by the Tevatron measurements ($\pm$~1.7~mb), and select three values for $\sigma_{eff}^{pp}$, namely, $\sigma_{eff}^{pp} = 13.3, 15$ and $16.7$~mb. The results can be seen in Fig.~\ref{pp7_eff}, showing a small deviation, approximately 10\%, regarding the central value $\sigma_{eff}^{pp} = 15$~mb. This implies in a weak dependence associated to the uncertainty on $\sigma_{eff}^{pp}$.

\begin{figure*}[!t]
\begin{tabular}{c}
\includegraphics[width=0.5\textwidth]{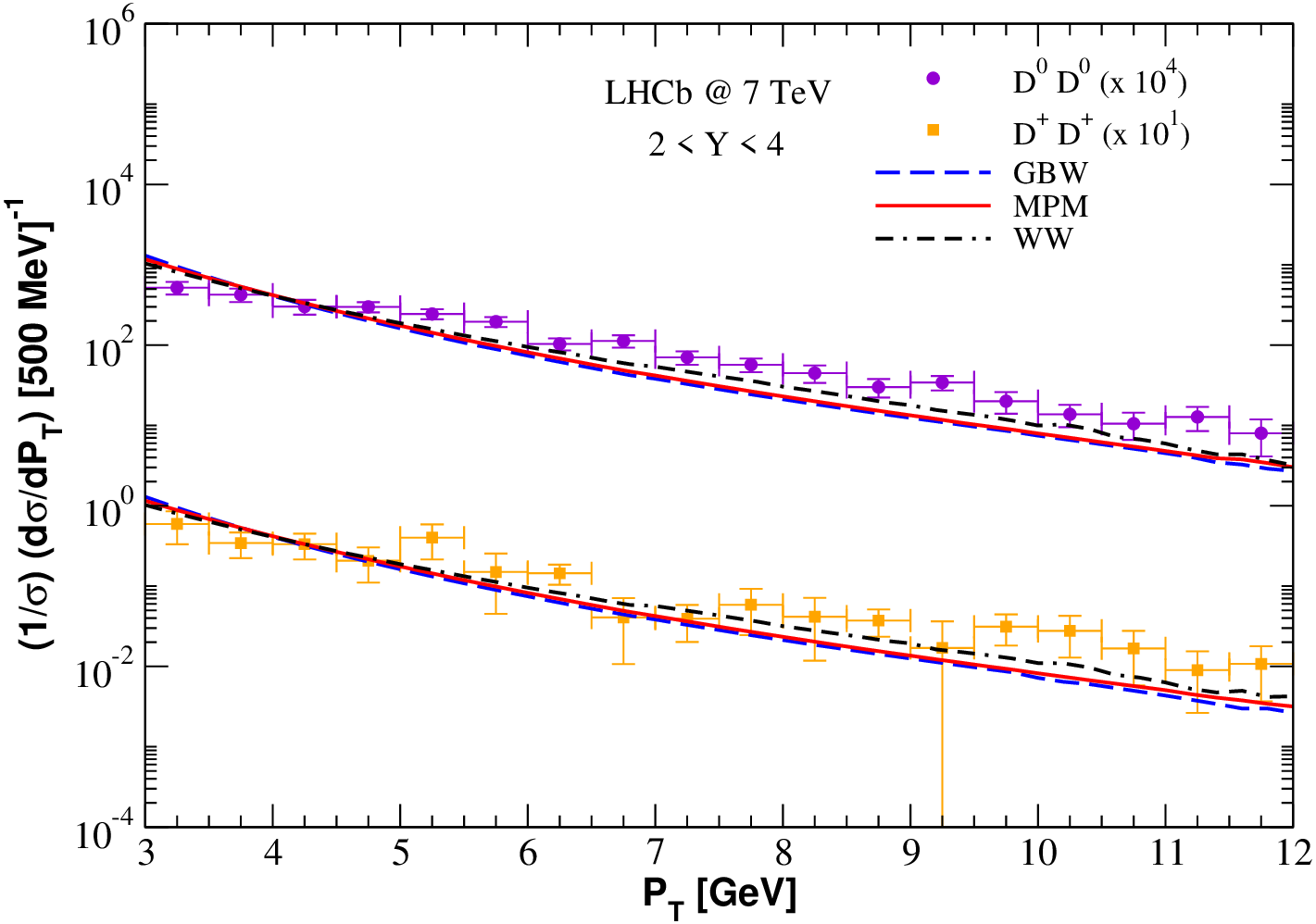} \includegraphics[width=0.5\textwidth]{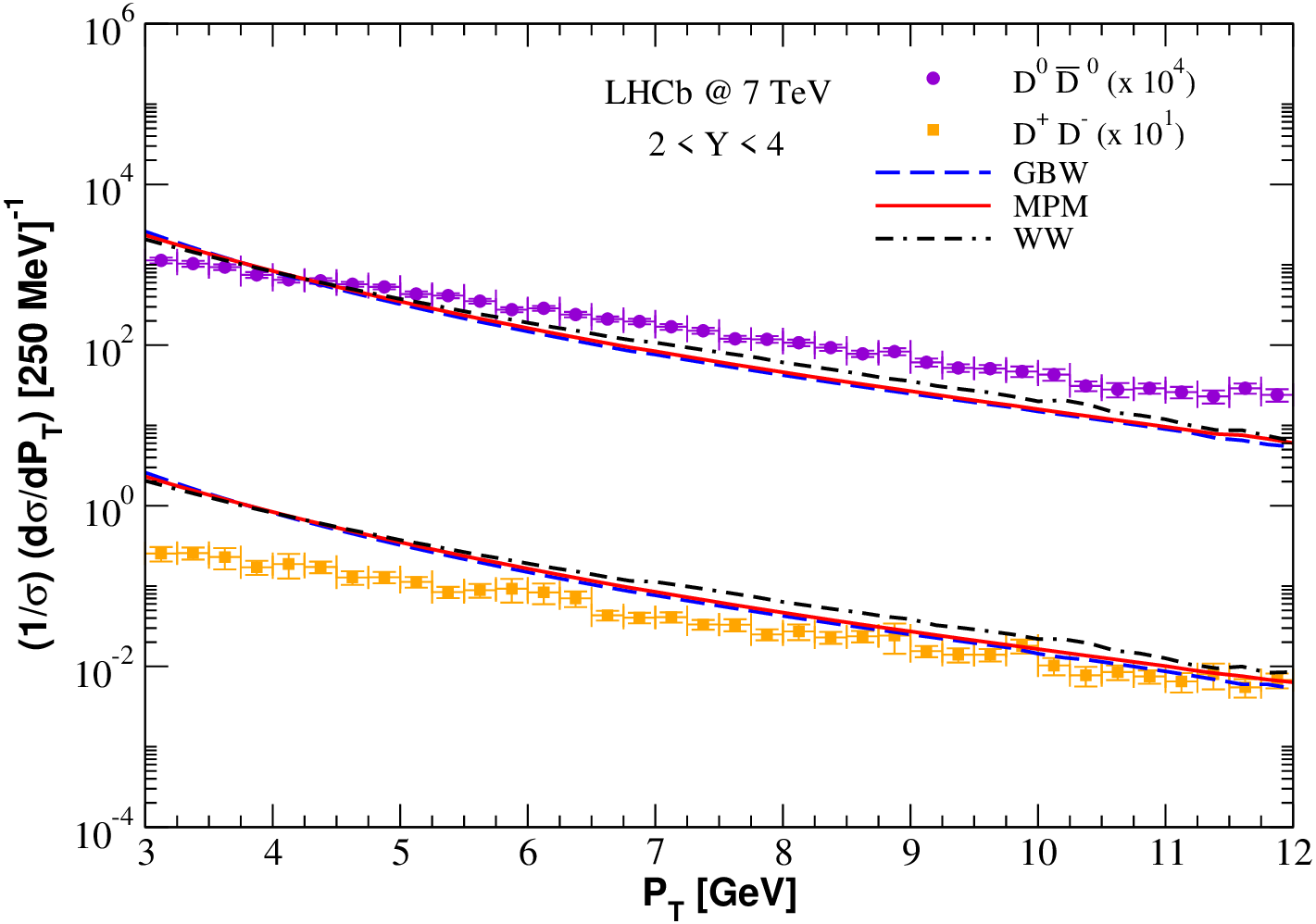}  \\
\includegraphics[width=0.5\textwidth]{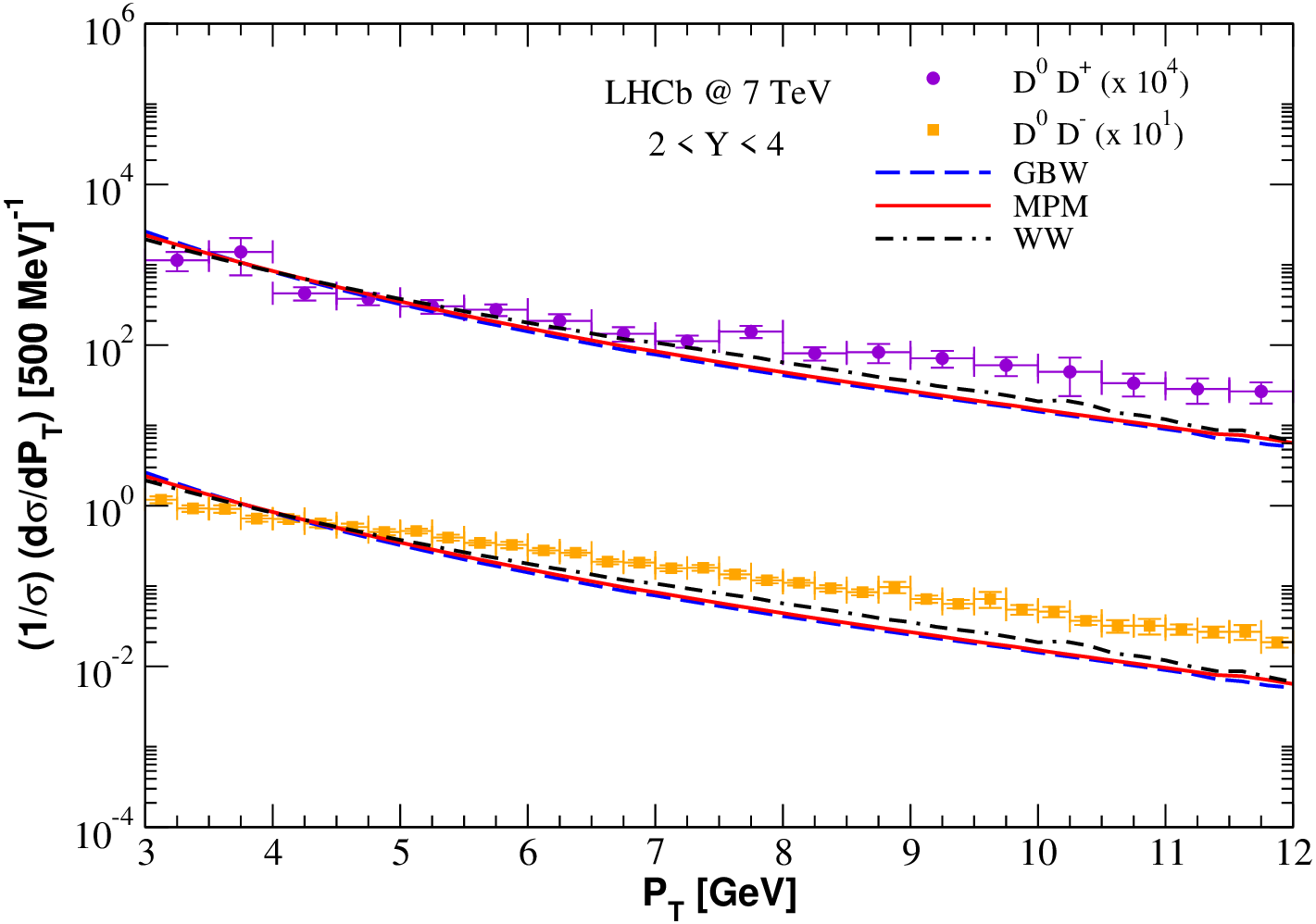}
\end{tabular}
\caption{Normalized differential cross section of the double charmed meson production in $pp$ collisions in terms of transverse momentum at $\sqrt{s} = 7$~TeV and $2 < Y < 4$ considering identical $D$-mesons (left panel), $D$-mesons and charge conjugate states (right panel), and two different $D$-mesons (bottom panel). The GBW, MPM, and WW predictions are compared with measurements provided by the LHCb \cite{LHCbpp}. The effective cross section is fixed as $\sigma_{eff}^{pp} = 15$~mb.}
\label{DD_pp7}
\end{figure*}

In the following we show our results for the $D$-meson pairs production cross sections in terms of the transverse momentum compared to the measurements performed by the LHCb experiment \cite{LHCbpp} in $pp$ collisions at $\sqrt{s} = 7$~TeV and for the rapidity bin $2 < Y < 4$. The Fig.~\ref{DD_pp7} displays the results assuming in the final state that the $D$-mesons are identical (left panel), the $D$-mesons and its corresponding charge conjugate states (right panel), and two different $D$-mesons (bottom panel), respectively. In all cases, the models fairly describe the experimental data for $P_T < 6$~GeV. Given the simplicity of the theoretical approach, the data description is reasonable bearing in mind that the normalization has been fixed by a particular choice of the effective cross section $\sigma_{eff}$. However, one cannot discriminate the models as they produce very similar results. A more pronounced deviation from data is observed towards large $P_T$ values, for instance, $P_T > 6$~GeV. In addition, some particular results are found considering the $D^{+}D^{+}$ and $D^{+}D^{-}$ results. In the former the data agreement on $P_T$ distribution is extended to $P_T = 9.5$~GeV, while for the latter the predictions overestimate the experimental points of the $P_T$ spectrum at $P_T < 8$~GeV, in contrast with a reasonable description of the data at the range 8~GeV~$< P_T < 12$~GeV.

Our predictions can be compared to other studies in literature. Investigation of DPS within the $k_{\perp}$-factorization formalism can be found in Ref.~\cite{maciula1}. There the authors discuss the $D^0D^0$ and $D^0\bar{D}^0$ production in $pp$ scattering assuming a double gluon fragmentation mechanism as well as the mixed gluon and charm DPS contribution. One consequence of these mechanisms is that a larger effective cross section was needed in order to describe the corresponding data. The value for $\sigma_{eff}^{pp} = 30$~mb is twice the usual values considered in DPS analyses. Interestingly, the channel $cc\rightarrow D^0D^0$ is subdominant in such approach whereas the $gg\rightarrow D^0D^0$ is the dominant one. A clear consequence of using several channels is the modification of the $p_T$-slope. It should be noticed that our calculations are fully consistent with those using $k_{\perp}$-factorization when only $cc\rightarrow D^0D^0$ is considered, as shown in Refs.~\cite{maciula2,vanHameren,vanHameren1}.

Along similar lines, in Ref.~\cite{Karpishkov:2016hnx} the double-$D$ production was addressed in the context of parton Reggeization approach. In this case, the hypothesis of double parton scattering is not involved and predictions are obtained without free parameters. There, the leading contribution to $D\bar{D}$ production is $gg$ fusion into charm pair with $c$ fragmentation into the $D$ meson and with $\bar{c}$ fragmentation into the $\bar{D}$ followed by the contribution from $gg$ fusion into two gluons which fragment into mesons. On the other hand, production of $DD$ pairs is mainly due to the gluon fragmentation into the $D$ meson in the subprocess of $gg$ fusion. Our calculations did not include the gluon fragmentation contributions and are limited by the DPS approximation given by the pocket formula.

In Ref.~\cite{Martinez:2018tuf} the double-$D$ inclusive production has been investigated in the CGC framework. The formalism includes both the production of two $c\bar{c}$ pairs as well as the production of one $c\bar{c}$ pair and a gluon. Unfortunately, the corresponding phenomenology has not been presented. Similarly to the parton Reggeization approach, the hypothesis of DPS is not implicated in the calculations. Of course, the DPS limit could be achieved by imposing uncorrelated initial partons in the framework.

\begin{figure*}[!t]
\begin{tabular}{c}
\includegraphics[width=0.75\textwidth]{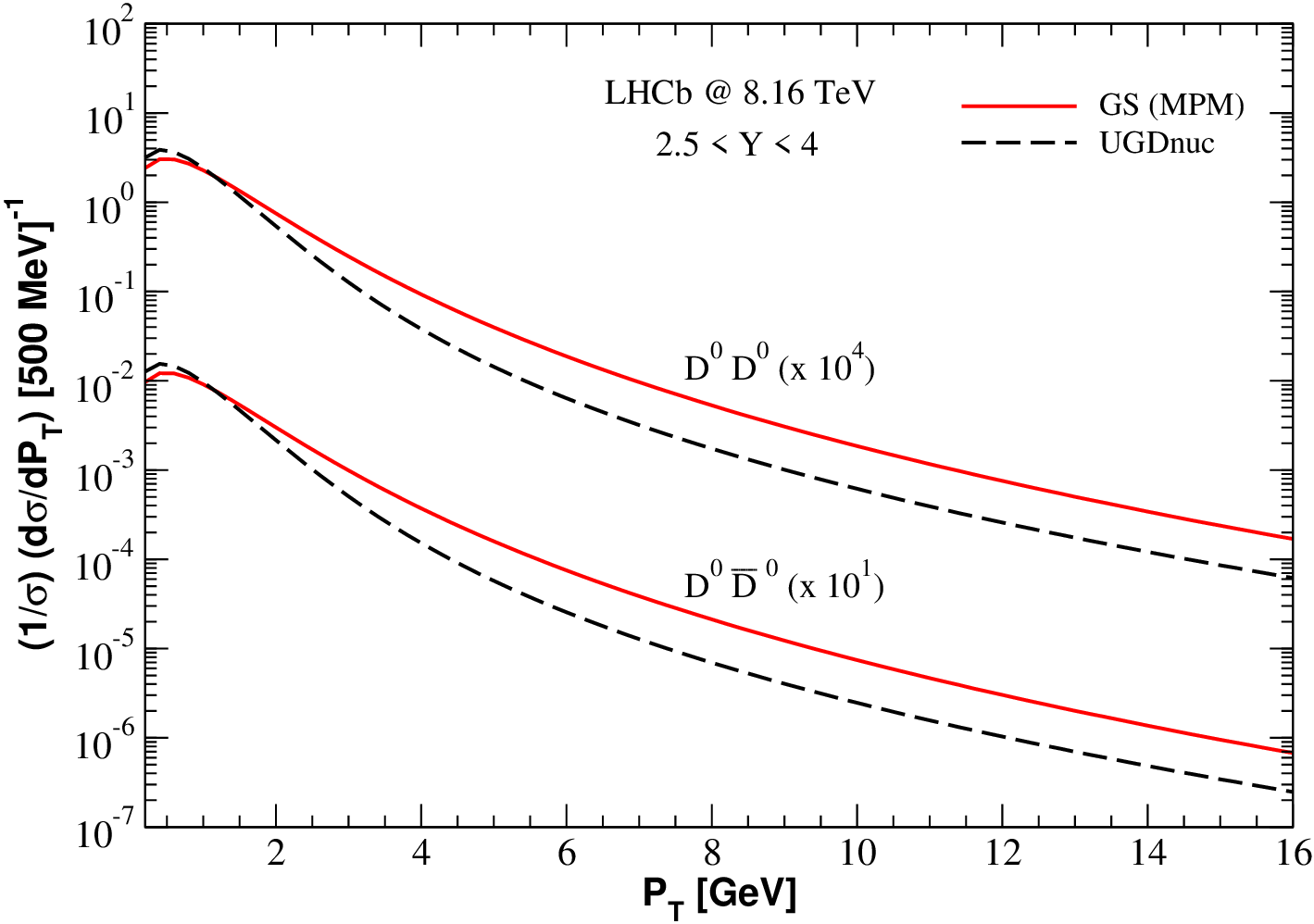} 
\end{tabular}
\caption{Normalized differential cross section of $D^0D^0$ and $D^0\bar{D}^0$ pairs production as function of transver momentum in $p$Pb collisions at $\sqrt{s} = 8.16$~TeV for $2 < Y < 4$. The predictions are obtained with \emph{GS (MPM)} and \emph{UGDnuc} models and an effective cross section $\sigma_{eff}^{pA}=4.3$~b.}
\label{DD_pPb816}
\end{figure*}

Now, in Fig.~\ref{DD_pPb816} we present the predictions for $D^0D^0$ and $D^0\bar{D}^0$ pairs production in $p$Pb collisions by means of the differential cross section as a function of $P_T$. The theoretical predictions consider the kinematic region that can be probed by the LHCb experiment defined by $\sqrt{s} = 8.16$~TeV and rapidity interval $2 < Y < 4$. Here, as mentioned before Glauber model was not considered to obtain the SPS $pA$ cross section.

The DPS cross section is computed by using Eq.~(\ref{dpsxs}), where $\sigma_{eff}^{pA}\approx\sigma_{inel}^{pA}\simeq 2$~b. This result is similar to that extracted by the LHCb Collaboration, where the value at forward rapidities \cite{LHCbpPb} for double-$D$ cross section is $1.41\pm 0.11 \pm 0.10$~b in the rapidity bin of $1.5<Y<4$. For simplicity, we employ the approximation $\sigma_{eff}^{pA}\approx A\times\sigma_{eff}^{pp}=4.3$~b at 8.16~TeV in our numerical calculations\footnote{The value 4.3~b is obtained by the $pp$ extrapolation performed in Ref.~\cite{LHCbpPb} by LHCb at 8.16~TeV which provides the $\sigma_{eff}^{pp}$ scaled by $A=208$ valid under the assumption of SPS production and no nuclear modification.}. The justification for such a simplification comes from the simplest DPS case, where the probability to produce particles $a$ and $b$ in a $pA$
collision is given as follows, $P_{pA\rightarrow ab} = P_{pA\rightarrow a} P_{pA\rightarrow b} =(\sigma_{pA\rightarrow a}/\sigma_{inel}^{pA})(\sigma_{pA\rightarrow b}/\sigma_{inel}^{pA})$, one leads to $\sigma_{pA\rightarrow ab} \simeq \sigma_{pA\rightarrow a}\sigma_{pA\rightarrow b}/\sigma_{inel}^{pA}$, with $\sigma_{eff} \approx \sigma_{inel}^{pA}$. By using the measured value of the inelastic cross section of 2061~mb at 5.02~TeV \cite{CMS:2015nfb} and the corresponding prediction for $pp$ collisions of 70~mb, one gets $\sigma_{inel}^{pA}\sim A^{0.63}\sigma_{inel}^{pp}$. Hence, we have assumed that the effective cross section roughly scales with $A$. As a matter of comparison, in the approach by d'Enterria and Snigirev (dES) \cite{dEnterria:2012jam} a sophisticated estimate is performed for the DPS $pA$ cross section. This is given by the sum of the two terms: (i) DPS cross section in pp collisions multiplied by A, and (ii) new contribution, for which interactions with partons from two different nucleons are involved in the scattering, related to the square of the thickness function (this contribution is not included in our case). Here, we should be careful: Eq.~(1) in our paper is different from Eq.~(15) in dES paper, as the numerator of Eq.~(1) involves the $p$A cross sections and not $pN$ ones. Assuming the hard scattering $\sigma_{pA}\approx A\sigma_{pN}$ in Eq.~(1), it is clear that $\sigma_{eff}^{pA}[\mathrm{ours}]\approx A^2\times \sigma_{eff,pA}[\mathrm{dES}]$. Therefore, this leads to $\sigma_{eff}^{pA}=A\times \sigma_{eff}^{pp}/[1+\sigma_{eff}^{pp}F_{pA}]\approx (A/3)\times \sigma_{eff}^{pp} $ at LHC energies for the dES approach in our notation. This is one third of the naive estimate considered in our work and is consistent with our approach.

The results differ very slightly in the small $P_T$ region, specifically for $P_T < 2$~GeV. One is able to verify that, aside from this specific kinematic domain, the models begin to provide distinct behaviors that become significant in the direction of large $P_T$ values. The \emph{GS (MPM)} predictions give a larger cross section than the \emph{UGDnuc} model, with deviations reaching 30\%. Moreover, our results can be used to discriminate between the approaches in view of data analysis of future experimental measurements of double $D$-meson production in $p$Pb collisions. 

Still about investigations of double charmed meson production in $p$Pb collisions, in Ref.~\cite{helenius} the calculations are based on the collinear factorized QCD at next-to-leading order using parton distribution functions and $D$-meson fragmentation functions. The authors provide results for the integrated cross section and for the projection of the relative azimuthal-angle distribution considering the $D^0D^0$ and $D^0\bar{D}^0$ pairs. In particular, the predictions assume a variation for $\sigma_{eff}^{pp}$, 10~mb~$< \sigma_{eff}^{pp} < 25$~mb, which is roughly the range deduced from jet, $W^{\pm}$ and photon measurements \cite{ATLAS}. For the nuclear case they used $\sigma_{eff}^{pPb}\simeq A\times\sigma_{eff}^{pp}/(2.5 \,...\, 4.8)$. We set the values in our estimates as 
$\sigma_{eff}^{pp} = 15$~mb and $\sigma_{eff}^{pPb} = 4.3$~b that are in relative agreement with the values extracted from the LHC measurements.

It is important to stress some aspects regarding the parameter $\sigma_{eff}$ that enters in the calculations. Commonly, the $\sigma_{eff}$ is determined by fitting the experimental measurements in order to be consistent with the corresponding data. Hence, there is an uncertainty associated to its value that may depend or not on the final state, which has been extracted \cite{calucci, treleani, ryskin, ATLAS}. Usually, $\sigma_{eff}$ relies on the kinematic variables related to the process. However, assuming the approximation that does not account the correlations between partons in the hadron \cite{david}, $\sigma_{eff}$ can be interpreted as a geometric quantity, establishing the pocket formula. Although the Eq.~(\ref{dpsxs}) has a factorized form derived in a simple baseline approach, phenomenological studies taking different observables into account show that the pocket formula can be successfully applied \cite{luszczak,cazaroto,cazaroto1,maciula1,helenius,berezhnoy,carvalho,maciula2,vanHameren}.
We also find in the literature investigations about correlations between the partons and double parton distributions in order to provide theoretical predictions. For example, in Ref.~\cite{gaunt} the authors provide a new set of DPDFs based on the LO DGLAP equation, where they derive momentum and number sum rules that the DPDFs must satisfy. There, they describe a program which uses a direct $x$-space method to numerically integrate the LO DGLAP equation and is used to evolve the input DPDFs to any other scale. The application of these DPDFs to the calculation of double-meson production is still unavailable. Accordingly, the pocket formula is recovered in case the longitudinal component $D^{ij}_h(x_1,x_2,Q_1,Q_2)$ of the DPDFs\footnote{The double parton distribution functions (DPDFs) are denoted by $\Gamma_{ab}(x_1,x_2,\vec{b}_1,\vec{b}_2,\mu_1^2,\mu_2^2)$ depend on the longitudinal momentum fractions $x_1$ and $x_2$ and on the transverse positions $\vec{b}_1$ and $\vec{b}_2$ of the two partons $a$ and $b$ undergoing the hard processes at the scales $\mu_1$ and $\mu_2$. Very often, it is assumed that the DPDFs may be decomposed in terms of the longitudinal, $D_h$, and transverse, $F_{\perp}$, components in the form  $\Gamma_{ab} =D_h^{ab}(x_1,x_2,\mu_1^2,\mu_2^2)F_{\perp}(\vec{b}_1,\vec{b}_2)$. The transverse part is given by $F_{\perp}(\vec{b}_1,\vec{b}_2) = f(\vec{b}_1)f(\vec{b}_2)$, where $f(\vec{b})$ is assumed as an universal function for all types of partons properly normalized.} is reduced to a product of two independent single parton distributions probed at resolution scales $Q_1$ and $Q_2$, respectively. The presence of the correlation term in the DPDFs results in the decrease of the effective cross section, $\sigma_{eff}$, with the growth of the resolution scales, while its dependence on the total energy at fixed scales is weaker  \cite{Snigirev:2010tk,Flensburg:2011kj,Ryskin:2011kk}.

As a last consideration, we discuss the validity range of the predictions using the QCD color dipole approach. The formalism would be applicable for $x_2\leq 0.01$ given that the parameters of the dipole cross section/UGD models are fitted to DIS data at Bjorken-$x$ $\leq 0.01$. Thus, it is convenient to investigate the $\langle x_2 \rangle$ value probed at the kinematic range analyzed by the LHCb measurements. For $pp$ collisions one has $\langle x_2 \rangle \sim 1 \times10^{-4}$, which gets a slightly smaller value of $\langle x_2 \rangle \sim 9 \times10^{-5}$ for $p$Pb collisions. One possible shortcoming is a possibly high $x_1$ value relative to the gluon distribution in the projectile in case of large $p_T$. In order to circumvent such limitation the GBW and MPM UGDs have been multiplied by a factor $(1-x_1)^7$. Therefore, the formalism is suitable for studying the $D$-meson pairs production in $pp/pA$ collisions at high energies.

\section{Summary} 
\label{conc}

We investigated the simultaneous $D$-meson pairs production in $pp$ and $pA$ collisions based on the color dipole framework in transverse momentum description using different unintegrated gluon distributions and considering double parton scattering mechanism. For $pA$ collisions we apply the geometric scaling property for the dipole-nucleus amplitude as well as a parameterization for the nuclear unintegrated gluon distribution in the Glauber-Gribov formalism. 

We demonstrated that the DPS processes need to be accounted in order to properly analyze $D$-meson double production, consequently, DPS contribution is essential to obtain the cross section being a substantial part of it. Particularly, in $pp$ collisions, the models GBW, MPM, and WW provide similar results that fairly describe the spectrum at $P_T < 6$~GeV, and they start to lose adherence to the spectrum from $P_T$ values above of $6$~GeV, where a deviation between the predictions begins to be visible.

Our results with \emph{GS (MPM)} and \emph{UGDnuc} for $p$Pb collisions present a difference in magnitude and they may be used to constrain the models using the momentum distribution associated to the differential cross section. This suggests that the corresponding future experimental measurements on the double $D$-meson production is feasible and that the analysis helps to probe the appropriate approach and its underlying assumptions. 

\section*{Acknowledgements}

We are grateful to Vanya Belyaev (ITEP Moscow) for valuable feedback on the LHCb data for double $D$-meson production in $pp$ collisions. This work was partially financed by the Brazilian funding agencies CAPES, CNPq, and FAPERGS. GGS
acknowledges funding from the Brazilian agency Conselho Nacional de Desenvolvimento Cient\'ifico e Tecnol\'ogico (CNPq) with grant CNPq/311851/2020-7.

\end{document}